\documentclass{ws-ijgmmp}

\usepackage[T1]{fontenc}
\usepackage[utf8]{inputenc}
\usepackage{amsopn,amstext,amssymb}
\usepackage[all]{xy}
\usepackage[english]{babel}
\usepackage{hyperref}

\let\tr\relax 

\newcommand\varnotation[1]{{\mathcal{#1}}}

\newcommand\algnotation[1]{{\mathbf{#1}}}

\newcommand\grnotation[1]{{\mathsf{#1}}}

\newcommand\modnotation[1]{{\boldsymbol{#1}}}

\newcommand\varE{{\varnotation{E}}}

\newcommand\varM{{\varnotation{M}}}

\newcommand\varP{{\varnotation{P}}}

\newcommand\algA{{\algnotation{A}}}
\newcommand\algB{{\algnotation{B}}}

\newcommand\algzero{{\grnotation{0}}}

\newcommand\modM{{\modnotation{M}}}

\newcommand\bbbone{{ \mathchoice {1\mskip-4mu\mathrm{l} } {1\mskip-4mu\mathrm{l} }{1\mskip-4.5mu\mathrm{l} } {1\mskip-5mu\mathrm{l}} }}

\newcommand\gR{{\mathbb R}}

\newcommand\gC{{\mathbb C}}

\newcommand\gN{{\mathbb N}}

\newcommand\caI{{\mathcal I}}

\newcommand\caL{{\mathcal L}}

\newcommand\caR{{\mathcal R}}
\newcommand\caS{{\mathcal S}}

\newcommand\caZ{{\mathcal Z}}

\newcommand\kg{{\mathfrak g}}

\newcommand\kS{{\mathfrak S}}

\newcommand\kX{{\mathfrak X}}
\newcommand\kY{{\mathfrak Y}}

\newcommand\ksl{{\mathfrak{sl}}}

\newcommand\ksu{{\mathfrak{su}}}

\newcommand\kisp{{\mathfrak{isp}}}

\newcommand{\omi}[1]{\buildrel { \buildrel{#1}\over{\vee} } \over .}
\newcommand{\spmatrix}[1]{\left( \begin{smallmatrix}#1\end{smallmatrix}\right)} 

\newcommand\exter{{\textstyle\bigwedge}} 

\newcommand\der{{\text{\textup{Der}}}}

\newcommand\Int{{\text{\textup{Int}}}}

\newcommand\Out{{\text{\textup{Out}}}}

\newcommand\sign{{\text{\textup{sign}}}}

\newcommand{\grast}{\bullet}

\newcommand{\moyast}{{\mathord{\star}}}

\newcommand\cdotaction{\mathord{\cdot}}

\DeclareMathOperator{\End}{End} 
\DeclareMathOperator{\Hom}{\mathsf{Hom}} 
\DeclareMathOperator{\tr}{Tr} 

\newcommand\adrep{{\text{\textup{ad}}}} 

\usepackage{lmodern}

\newcommand\dd{\text{\textup{d}}}
\newcommand\hd{\widehat{\dd}}
\newcommand\hR{\widehat{R}}
\newcommand{\fibre}[4][]{\ensuremath{\xymatrix@1@C=14pt{{#2} \ar[r] & {#3} \ar[r]^-{#1} & {#4}}}}

\begin{document}

\markboth{Examples of derivation-based differential calculi related to noncommutative gauge theories}{Examples of derivation-based differential calculi related to noncommutative gauge theories}

%
\catchline{}{}{}{}{}
%

\title{Examples of derivation-based differential calculi related to noncommutative gauge theories}
\author{T. Masson}
\address{Laboratoire de Physique Th\'eorique (UMR 8627)\\
B\^at 210, Universit\'e Paris-Sud Orsay\\
F-91405 Orsay Cedex}

\maketitle

\begin{abstract}
Some derivation-based differential calculi which have been used to construct models of noncommutative gauge theories are presented and commented. Some comparisons between them are made.
\end{abstract}


\begin{flushleft}
LPT-Orsay/08-85
\end{flushleft}


\begin{flushright}
\slshape à Michel,\\ 
qui m'a donné le goût\\
de la simplicité et de la rigueur.
\end{flushright}

\section{Introduction}

There are many approaches to noncommutative geometry. Some of them should be mentioned by the more accurate terminology of ``noncommutative riemannian geometry'', while some should be referred to as ``noncommutative differential geometry''. The point is that there is some confusion on what the generic denomination ``noncommutative geometry'' means, and what it is for.

The spectral triple approach by A. Connes is obviously in the prototype of noncommutative riemannian geometry, because it emphases on the metric structure before the differential one. The notion of spectral triple has been forged on the commutative model of spin geometry, where riemannian metrics play an essential role.

On the other hand, some approaches have focused on the differential objects one can construct on noncommutative algebras. Very often, this noncommutative ``differential'' geometry is encoded into a differential calculus. It is well known that any associative algebra can be equipped with an universal differential calculus (see \cite{Mass:31} for the notion of differential calculus in general, and the universal one in particular). But this differential calculus does not contain much of what one would like to be a ``noncommutative geometry'', so that definitions of differential calculi have been proposed in different contexts, regarding properties of the algebras under considerations.

For instance, for quantum groups, the notion of differential calculus has been introduced through the requirement of covariance (see the monographs \cite{CharPres:94} and \cite{KlimSchm:97} for details).

For general associative algebras, it is well known that the infinitesimal version of automorphisms are derivations. This is the starting point of the derivation-based differential calculus introduced by M.~Dubois-Volette in \cite{DuVi:88} and described in this paper.

These noncommutative differential geometries have been applied to different situations. For instance, attempts have been made to describe noncommutative linear connections in a large class of noncommutative geometries (\cite{Mass:03, Mass:04, Mass:05, Mass:06, Mass:10, Mass:12} for general considerations and concrete examples and \cite{Mado:99a} for a review). 

But the truly interest in these noncommutative differential geometries relies on their applications to gauge field theories. Indeed, a pioneer result by M.~Dubois-Violette, R.~Kerner and J.~Madore in \cite{DuViKernMado:90b} showed that the derivation-based noncommutative calculus can be used to generate some Yang-Mills-Higgs models elegantly, in which the Higgs fields are the components in the purely noncommutative directions of noncommutative connections. Some other differential calculi share this very interesting feature, as the one introduced in \cite{Connes:1990qp} and further developed to a full version of the Standard Model of elementary particles in \cite{Chamseddine:2006ep}. 

Notice that methods of differential algebra, through BRST methods, has already been applied successfully to master symmetries of (ordinary) gauge theories, see for instance \cite{Becchi:1975nq}, \cite{Ader:1987kh} and \cite{DuVi:86}. The present framework, especially \cite{DuViKernMado:90b}, reveals surprising similarities and relationship with these BRST considerations, whose study has not yet been pursed.

In this paper, I focus entirely on the derivation-based differential calculus, for which I give some examples that are related to gauge theories and this Higgs mechanism.

\section{The derivation-based differential calculus}
\label{Thederivationbaseddifferentialcalculus}

The notion of derivation-based differential calculus was introduced by M.~Dubois-Volette in \cite{DuVi:88} and further studied in \cite{DuViMich:94} and \cite{DuViMich:96}. In \cite{DuVi:01a}, a more systematic approach is proposed through the use of the categorical point of view on algebras.
 
The idea was to introduce a general and purely algebraic definition of a differential calculus associated to any (commutative or non commutative) associative algebra, which mimics the construction of the de~Rham forms on the algebra $C^\infty(\varM)$ of smooth functions on a smooth paracompact manifold.

In order to do that, one has to notice that there is an algebraic definition of the de~Rham complex, based on the notion of derivations of the algebra $C^\infty(\varM)$. This construction can be directly used in the non commutative case, except for some details concerning the role of the center of the associative algebra.

The potential of using the space of derivations to study some aspects of the ``non commutative geometry'' of some algebras has been highlighted by A.~Connes in \cite{Conn:85} where he computed the Hochschild and cyclic cohomology of the irrational rotation algebra (the non commutative torus at irrational values of the deformation parameter). In the characterization of these cohomologies, the two derivations of this algebra play a fundamental role. Notice that the noncommutative torus plays an interesting role in the so called topological description of the quantum Hall effect, which supplements other more physical approaches based, for instance, on hierarchical constructions (see e.g. \cite{GeorWall:97a}, \cite{Mass:13} and \cite{Mass:17} and references therein).

The systematic use of the space of derivation of an associative algebra as a candidate for the study of its non commutative geometry was initiated by M.~Dubois-Violette in \cite{DuVi:88}, where he introduced and studies the so-called derivation-based differential calculus. 
Many examples of this differential calculus have been considered so far, and only some of them will be considered here. The choice I have made rely on the declination of two fundamental examples I will introduced in the next section: the algebra of smooth functions and the (finite dimensional) matrix algebra. The other examples I will expose are constructed using these ``building blocks'' and they will be compared to them.

Because of this choice, many other interesting examples of derivation-based differential calculi will not be exposed here, and I refer to \cite{DuVi:91}, \cite{DuViMich:94} and \cite{DuViMich:96} for further developments. In particular, the example of the canonical commutation relations algebra exposed in \cite{DuVi:91} is very suggestive concerning the deep relations between classical and quantum mechanics.

\subsection{The differential calculus}

In the following, $\algA$ denotes an associative algebra with unit $\bbbone$, and $\caZ(\algA)$ denotes its center.

\begin{definition}[The vector space of derivations of $\algA$]
The vector space of derivations of $\algA$ is the space defined by\\
$\der(\algA) = \{ \kX : \algA \rightarrow \algA \ / \ \kX \text{ linear}, \kX(ab) = \kX(a) b + a \kX(b), \forall a,b\in \algA\}$
\end{definition} 

The construction of the differential calculus based on derivations relies on the following properties of the space of derivations.

\begin{proposition}[Structure of $\der(\algA)$]
$\der(\algA)$ is a Lie algebra for the bracket $[\kX, \kY ]a = \kX  \kY a - \kY \kX a$ ($\forall \kX,\kY \in \der(\algA)$) and a $\caZ(\algA)$-module for the product $(f\kX )a = f(\kX a)$ ($\forall f \in \caZ(\algA)$, $\forall \kX \in \der(\algA)$).

The subspace $\Int(\algA) = \{ \adrep_a : b \mapsto [a,b]\ / \ a \in \algA\} \subset \der(\algA)$, called the vector space of inner derivations, is a Lie ideal and a $\caZ(\algA)$-submodule.

With $\Out(\algA)=\der(\algA)/\Int(\algA)$, there is a short exact sequence of Lie algebras and $\caZ(\algA)$-modules
\begin{equation}
\label{eq-secderivations}
\xymatrix@1@C=15pt{{\algzero} \ar[r] & {\Int(\algA)} \ar[r] & {\der(\algA)} \ar[r] & {\Out(\algA)} \ar[r] & {\algzero}}
\end{equation}
\end{proposition} 

If the algebra $\algA$ is equipped with an involution $a \mapsto a^\ast$, one can define real derivations as follows:

\begin{definition}[Real derivations of an involutive algebra]
If $\algA$ is an involutive algebra, the derivation $\kX \in \der(\algA)$ is real if $(\kX a)^\ast = \kX a^\ast$ for any $a\in \algA$. The space of real derivations will be denoted by $\der_\gR(\algA)$.
\end{definition} 

Using the properties of the space $\der(\algA)$, one can introduce the following graded differential algebra:

\begin{definition}[The graded differential algebra $\underline{\Omega}^\grast_\der(\algA)$]
Let $\underline{\Omega}^n_\der(\algA)$ be the set of $\caZ(\algA)$-multilinear antisymmetric maps from $\der(\algA)^n$ to $\algA$, with $\underline{\Omega}^0_\der(\algA) = \algA$, and let
\begin{equation*}
\underline{\Omega}^\grast_\der(\algA) =\textstyle \bigoplus_{n \geq 0} \underline{\Omega}^n_\der(\algA)
\end{equation*}
The space $\underline{\Omega}^\grast_\der(\algA)$ can be equipped with a structure of $\gN$-graded differential algebra using the product
\begin{multline*}
(\omega\eta)(\kX_1, \dots, \kX_{p+q}) = \\
 \frac{1}{p!q!} \sum_{\sigma\in \kS_{p+q}} (-1)^{\sign(\sigma)} \omega(\kX_{\sigma(1)}, \dots, \kX_{\sigma(p)}) \eta(\kX_{\sigma(p+1)}, \dots, \kX_{\sigma(p+q)})
\end{multline*}
and the differential $\dd$ (of degree $1$) defined by the Koszul formula
\begin{multline*}
\dd\omega(\kX_1, \dots , \kX_{n+1}) = \sum_{i=1}^{n+1} (-1)^{i+1} \kX_i \omega( \kX_1, \dots \omi{i} \dots, \kX_{n+1}) \\[-5pt]
 + \sum_{1\leq i < j \leq n+1} (-1)^{i+j} \omega( [\kX_i, \kX_j], \dots \omi{i} \dots \omi{j} \dots , \kX_{n+1}) 
\end{multline*}
\end{definition} 

This differential graded algebra contains a particular graded differential subalgebra:

\begin{definition}[The graded differential algebra $\Omega^\grast_\der(\algA)$]
We denote by $\Omega^\grast_\der(\algA) \subset \underline{\Omega}^\grast_\der(\algA)$ the differential graded subalgebra of $\underline{\Omega}^\grast_\der(\algA)$ generated in degree $0$ by $\algA$.
\end{definition} 

By construction, every element in $\Omega^n_\der(\algA)$ is a sum of terms of the form $a_0 \dd a_1 \cdots \dd a_n$ for $a_0, \dots, a_n \in \algA$. This is not necessary the case for any element in $\underline{\Omega}^\grast_\der(\algA)$. Nevertheless, in the examples we will encounter, these two spaces will coincide.

\bigskip
The previous definitions were motivated by the following important (commutative) example which shows why these definitions are correct generalisations of the space of ordinary differential de~Rham forms on a manifold:
\begin{example}[The algebra $\algA = C^\infty(\varM)$]
\label{example-Cinfty}
Let $\varM$ be a smooth paracompact manifold and let $\algA = C^\infty(\varM)$ the commutative algebra of smooth functions on $\varM$. Obviously, the center of this algebra is $\algA$ itself: $\caZ(\algA) = C^\infty(\varM)$. Using well known results in differential geometry, the Lie algebra of derivations is exactly the Lie algebra of smooth vector fields on $\varM$: $\der(\algA) = \Gamma(\varM)$. In that case, there is no inner derivations, $\Int(\algA) = \algzero$, so that $\Out(\algA) = \Gamma(\varM)$.

The two graded differential algebras coincide with the graded differential algebra of de~Rham forms on $\varM$: $\Omega^\grast_\der(\algA) = \underline{\Omega}^\grast_\der(\algA) = \Omega^\grast(\varM)$.
\end{example}

This is a fundamental illustration of these definitions, and we will see in the next section how this construction make sense also in some non commutative situations.

\bigskip
In the previous definitions of the graded differential calculi, one is not bounded to consider the full Lie algebra of derivations:
\begin{definition}[Restricted derivation-based differential calculus]
Let $\kg \subset \der(\algA)$ be a sub Lie algebra and a sub $\caZ(\algA)$-module. The restricted derivation-based differential calculus $\underline{\Omega}^\grast_\kg(\algA)$ associated to $\kg$ is defined as the set of $\caZ(\algA)$-multilinear antisymmetric maps from $\kg^n$ to $\algA$ for $n \geq 0$, using the previous formulae for the product and the differential.
\end{definition} 

\bigskip
In the study of the geometry of fibre bundles through differential forms, the notion of Cartan operation is one of the main tools. This notion can be defined in the noncommutative geometry of the derivation-based differential calculus $(\underline{\Omega}^\grast_\der(\algA), \dd)$, using the following procedure. 
\begin{definition}[Cartan operation]
Let $\kg$ be any Lie subalgebra of $\der(\algA)$. The interior product is the graded derivation of degree $-1$ on $\underline{\Omega}^\grast_\der(\algA)$ defined by
\begin{align*}
i_\kX : \underline{\Omega}^n_\der(\algA)  &\rightarrow \underline{\Omega}^{n-1}_\der(\algA)
&
(i_\kX \omega)( \kX_1, \dots , \kX_{n-1})  = \omega (\kX, \kX_1, \dots , \kX_{n-1})
\end{align*}
$\forall \kX\in \kg$, $\forall \omega \in \underline{\Omega}^n_\der(\algA)$ and $ \forall \kX_i \in \der(\algA)$. By definition, $i_\kX$ is $0$ on $\underline{\Omega}^0_\der(\algA)=\algA$.

The associated Lie derivative is the graded derivation of degree $0$ on $\underline{\Omega}^\grast_\der(\algA)$ given by
\begin{equation*}
L_\kX = i_\kX \dd + \dd i_\kX  : \underline{\Omega}^n_\der(\algA) \rightarrow \underline{\Omega}^{n}_\der(\algA)
\end{equation*}

These graded derivations gives rise to a Cartan operation of $\kg$ on $(\underline{\Omega}^\grast_\der(\algA), \dd)$ in the sense that 
\begin{align*}
i_\kX i_\kY + i_\kY i_\kX &= 0 & L_\kX i_\kY - i_\kY L_\kX &= i_{[\kX,\kY]} \\
L_\kX L_\kY - L_\kY L_\kX &= L_{[\kX,\kY]} & L_\kX \dd - \dd L_\kX &= 0
\end{align*}
\end{definition}

As usual, one can associate to such an operation the following subspaces of $\underline{\Omega}^\grast_\der(\algA)$:
\begin{itemize}
\item The horizontal subspace is the kernel of all the $i_\kX$ for $\kX\in \kg$. This is a graded algebra.

\item The invariant subspace is the kernel of all the $L_\kX$ for $\kX\in \kg$. This is a graded differential algebra.

\item The basic subspace is the kernel of all the $i_\kX$ and $L_\kX$ for $\kX\in \kg$. This is a graded differential algebra.
\end{itemize}

For instance, $\kg = \Int(\algA)$ defines such an operation.

\subsection{Noncommutative connections}

The theory of connections is an essential ingredient in ordinary differential geometry and it is strongly connected to gauge fields theories in particle physics. This theory admits some noncommutative versions, depending essentially on the differential calculus one considers. In the case of the derivation-based differential calculus, noncommutative connections look very similar to ordinary connections, as it will be seen in the following. 

Every theory of noncommutative connections relies on modules over the associative algebra one wants to study. In general, one takes left or right finite projective modules. In the context of derivation-based differential calculus, it has been possible to consider bimodules. We refer to \cite{DuViMich:96} for some developements in this direction. Nevertheless, in the following, I will restrict myself to right modules, because in that case, the gauge group one can associate is big enough to give rise to interesting developments in physics.

In the following, $\modM$ is a right $\algA$-module.

\begin{definition}[Noncommutative connection, curvature]
A noncommutative connection on $\modM$ is a linear map $\widehat{\nabla}_\kX : \modM \rightarrow \modM$, defined for any $\kX \in \der(\algA)$, such that $\forall \kX,\kY \in \der(\algA)$, $\forall a \in \algA$, $\forall m \in \modM$, $\forall f \in \caZ(\algA)$:
\begin{align*}
\widehat{\nabla}_\kX (m a) &= m(\kX a) + (\widehat{\nabla}_\kX m) a,
&
\widehat{\nabla}_{f\kX} m &= f \widehat{\nabla}_\kX m,
&
\widehat{\nabla}_{\kX + \kY} m &= \widehat{\nabla}_\kX m + \widehat{\nabla}_\kY m
\end{align*}

The curvature of $\widehat{\nabla}$ is the linear map $\hR(\kX, \kY) : \modM \rightarrow \modM$ defined for any $\kX, \kY \in \der(\algA)$ by $\hR(\kX, \kY) m = [ \widehat{\nabla}_\kX, \widehat{\nabla}_\kY ] m - \widehat{\nabla}_{[\kX, \kY]}m$.
\end{definition} 

This definitions of $\widehat{\nabla}$ and $\hR$ are very similar to the one encountered in ordinary differential geometry, because of the presence of derivations in the expressions in place of vector fields. For other differential calculi, the lack of ``argument'' for noncommutative forms does not permit one to write down similar expression, so that one has to deal with maps $\widehat{\nabla} : \modM \rightarrow \modM \otimes_{\algA} \Omega^1$ and $\hR : \modM \rightarrow \modM \otimes_{\algA} \Omega^2$ where $\Omega^p$ is the space (bimodule) of $p$-forms.

\begin{proposition}[General properties]
The space of connections is an affine space modeled over the vector space $\Hom^{\algA}(\modM, \modM \otimes_\algA\underline{\Omega}^1_\der(\algA))$ (right $\algA$-module homomorphisms).

$\hR(\kX, \kY) : \modM \rightarrow \modM$ is a right $\algA$-module homomorphism for any $\kX,\kY \in \der(\algA)$.
\end{proposition} 

Notice that \textit{a priori} we do not know if $\kX \mapsto \widehat{\nabla}_\kX m$ is in $\modM \otimes_\algA\Omega^1_\der(\algA))$ so that one really needs to introduce the graded differential algebra of forms $(\underline{\Omega}^\grast_\der(\algA), \dd)$.

\begin{definition}[Gauge group]
The gauge group of $\modM$ is the group of automorphisms of $\modM$ as a right $\algA$-module.
\end{definition} 

This definition does not use any notion of connection, so that the gauge group is given directly from the module structure. But the following shows that this gauge group is compatible with the notion of connection:

\begin{proposition}[Gauge transformations]
For any $\Phi$ in the gauge group of $\modM$ and any noncommutative connection $\widehat{\nabla}$, the map $\widehat{\nabla}^\Phi_\kX = \Phi^{-1}\circ \widehat{\nabla}_\kX \circ \Phi : \modM \rightarrow \modM$ is a noncommutative connection.

This defines the action of the gauge group on the space of noncommutative connections.

This action induced an action on the curvatures of connections.
\end{proposition} 

\bigskip
Consider now the case of the right $\algA$-module $\modM=\algA$. In the following, $\algA$ is a unital algebra.

Let $\widehat{\nabla}_\kX : \algA \rightarrow \algA$ be a noncommutative connection.

\begin{proposition}[Noncommutative connections on $\modM=\algA$]
The connection $\widehat{\nabla}$ is completely given by it value $\widehat{\nabla}_\kX \bbbone = \omega(\kX)$, with $\omega \in \underline{\Omega}^1_\der(\algA)$, through the reconstruction relation 
\begin{equation*}
\widehat{\nabla}_\kX a = \kX a + \omega(\kX) a
\end{equation*}

The curvature of $\widehat{\nabla}$ is the multiplication on the left on $\algA$ by the noncommutative $2$-form given by
\begin{equation*}
\Omega(\kX, \kY) = \dd\omega (\kX, \kY) + [ \omega(\kX), \omega(\kY) ]
\end{equation*}

The gauge group is identified with the invertible elements $g \in \algA$ by $\Phi_g(a) = ga$.

The gauge transformations on $\widehat{\nabla}$ take the following form on the two noncommutative forms $\omega$ and $\Omega$:
\begin{align*}
\omega &\mapsto \omega^g = g^{-1} \omega g + g^{-1} \dd g
&
\Omega &\mapsto \Omega^g = g^{-1} \Omega g
\end{align*}
$\widehat{\nabla}^0_\kX$ defined by $a \mapsto \kX a$ is a noncommutative connection on $\algA$.
\end{proposition}

\section{Fundamental examples}

The first fundamental example I do not need to further comment is the algebra $\algA = C^\infty(\varM)$ of Example~\ref{example-Cinfty}. All the notions of noncommutative connections, curvature, gauge groups and gauge transformations coincide with the usual ones in this context.

\subsection{The matrix algebra $\algA = M_n(\gC)$}

The study of this example has been initiated in \cite{DuViKernMado:90a}. A complete description of this differential calculus can be found in \cite{Mass:30} and \cite{Mass:31}. This algebra is equipped with its usual involution.

The main results can be summarize in the following:
\begin{proposition}[General properties of the differential calculus]
\label{prop-differentialcalculusmatrixalgebra}
One has the following results:
\begin{itemize}
\item $\caZ(M_n) = \gC$.

\item $\der(M_n) = \Int(M_n) \simeq \ksl_n =\ksl(n,\gC)$ (traceless matrices). The explicit isomorphism associates to any $\gamma \in \ksl_n(\gC)$ the derivation $\adrep_\gamma : a \mapsto [\gamma, a]$.\\
$\der_\gR(M_n) = \ksu(n)$ and $\Out(M_n) = \algzero$.

\item $\underline{\Omega}^\grast_\der(M_n) = \Omega^\grast_\der(M_n) \simeq M_n \otimes \exter^\grast \ksl_n^\ast$, where $\dd'$ is the differential of the differential complex of the Lie algebra $\ksl_n$ represented on $M_n$ by the adjoint representation (commutator).

\item There exits a canonical noncommutative $1$-form $i\theta \in \Omega^1_\der(M_n)$ such that, for any $\gamma \in M_n(\gC)$,
\begin{equation*}
i\theta(\adrep_{\gamma}) = \textstyle\gamma - \frac{1}{n} \tr (\gamma)\bbbone
\end{equation*}
This noncommutative $1$-form $i\theta$ makes the explicit isomorphism $\Int(M_n(\gC)) \xrightarrow{\simeq} \ksl_n$.

\item $i\theta$ satisfies the relation $\dd' (i\theta) - (i\theta)^2 = 0$. This makes $i\theta$ look very much like the Maurer-Cartan form in the geometry of Lie groups (here $SL_n(\gC)$).

\item For any $a \in M_n$, one has $\dd' a = [i\theta, a] \in \Omega^1_\der(M_n)$. This relation is no longer true in higher degrees.
\end{itemize}

\end{proposition}

It is convenient to introduce a particular basis of this algebra. This permits one to perform explicit computations. Let $\{E_k\}_{k=1, \dots, n^2-1}$ be a basis for $\ksl_n$ of hermitean matrices. By the isomorphism of the previous proposition, it defines a basis for the Lie algebra $\der(M_n) \simeq \ksl_n$ through the $n^2-1$ derivations $\partial_k = ad_{i E_k}$, which are real derivations. Adjoining the unit $\bbbone$ to the $E_k$'s, one gets a basis for $M_n$.

Let the $\theta^\ell$'s be defined in $\ksl_n^\ast$ by duality: $\theta^\ell(\partial_k) = \delta^\ell_k$. Then $\{\theta^\ell\}$ is a basis of $1$-forms in $\exter^\grast \ksl_n^\ast$. By definition of the product of forms, they anticommute: $\theta^\ell \theta^k = - \theta^k \theta^\ell$.

Introducing the structure constants by $[E_k, E_\ell] = C^m_{k \ell} E_m$, the differential $\dd'$ takes the explicit form:
\begin{align*}
\dd' \bbbone &= 0 & \dd' E_k &= -C^m_{k\ell} E_m \theta^\ell & \dd' \theta^k &= -\frac{1}{2} C^k_{\ell m} \theta^\ell \theta^m
\end{align*}

The noncommutative $1$-form $i\theta$ can be written explicitly as $i\theta = i E_k \theta^k \in M_n \otimes \exter^1 \ksl_n^\ast$. This expression is obviously independent of the chosen basis.

\begin{proposition}[The cohomology of the differential calculus]
The cohomology of the differential algebra $(\Omega^\grast_\der(M_n), \dd')$ is
\begin{equation*}
H^\grast( \Omega^\grast_\der(M_n), d') = \caI(\exter^\grast\ksl_n^\ast)
\end{equation*}
the algebra of invariant elements for the natural Lie derivative. 

Recall that the algebra $\caI(\exter^\grast\ksl_n^\ast)$ is the graded commutative algebra generated by elements $c^n_{2 r - 1}$ in degree $2r-1$ for $r \in \{ 2,3, \dots, n\}$.
\end{proposition} 

This noncommutative geometry can be equipped with a metric. In order to do that, let us introduce the symmetric matrix $g_{k \ell} = \frac{1}{n} \tr(E_k E_\ell)$. Then the $g_{k \ell}$'s define a natural scalar product on $\der(M_n)$ by the relation $g(\partial_k, \partial_\ell) = g_{k \ell}$.

Moreover, any differential form of maximal degree $\omega \in \Omega^{n^2-1}_\der(M_n)$ can be written uniquely in the form
\begin{equation*}
\omega = a \sqrt{|g|} \theta^1 \cdots \theta^{n^2-1}
\end{equation*}
where $a \in M_n$ and where $|g|$ is the determinant of the matrix $(g_{k \ell})$. This gives rise to the following:

\begin{definition}[Noncommutative integration]
One defines a noncommutative integration 
\begin{equation*}
\int_{\text{n.c.}} : \Omega^\grast_\der(M_n) \rightarrow \gC
\end{equation*}
by $\int_{\text{n.c.}} \omega = \frac{1}{n} \tr(a)$ if $\omega \in \Omega^{n^2-1}_\der(M_n)$ written as above, and $0$ otherwise.

This integration satisfies the closure relation
\begin{equation*}
\int_{\text{n.c.}} \dd' \omega = 0
\end{equation*}
\end{definition}

The triple $(\Omega^\grast_\der(M_n), \dd', \int_{\text{n.c.}})$ is a cycle in the sense of Connes~\cite{Conn:85}.

\bigskip
Let us now take a brief look at gauge theories in this geometry. Consider first the right $\algA$-module $\modM = \algA$. 

The noncommutative $1$-form $-i\theta$ defines a canonical noncommutative connection by the relation $\widehat{\nabla}^{-i\theta}_\kX a = \kX a - i\theta(\kX) a$ for any $a \in \algA$. This is a common feature of such a $1$-form related to the differential in degree $0$ by a commutator (see~\cite{Mass:30}).

\begin{proposition}[Properties of $\widehat{\nabla}^{-i\theta}$]
For any $a \in M_n$ and $\kX = \adrep_\gamma \in \der(M_n)$ (with $\tr \gamma = 0$), one has 
\begin{equation*}
\widehat{\nabla}^{-i\theta}_\kX a = - a i\theta(\kX) = - a \gamma
\end{equation*}

$\widehat{\nabla}^{-i\theta}$ is gauge invariant and its curvature is zero.
\end{proposition}

Let us now consider the right $\algA$-module $\modM = M_{r,n}$, the vector space of $r\times n$ complex matrices with the obvious right module structure and the Hermitean structure $\langle m_1, m_2 \rangle = m_1^\ast m_2 \in M_n$.

\begin{proposition}[$\widehat{\nabla}^{-i\theta}$, flat noncommutative connections]
The noncommutative connection $\widehat{\nabla}^{-i\theta}_\kX m = - m i\theta(\kX)$ is well defined, it is compatible with the Hermitean structure and its curvature is zero.

Any noncommutative connection can be written as $\widehat{\nabla}_\kX a = \widehat{\nabla}^{-i\theta}_\kX a + A(\kX) a$ for $A = A_k \theta^k$ with $A_k \in M_r$. The curvature of $\widehat{\nabla}$ is the multiplication on the left by the $M_r$-valued noncommutative $2$-form
\begin{equation*}
F = \frac{1}{2}( [A_k, A_\ell] - C^m_{k \ell} A_m) \theta^k \theta^\ell
\end{equation*}
This curvature vanishes if and only if $A : \ksl_n \rightarrow M_r$ is a representation of the Lie algebra $\ksl_n$.

Two flat connections are in the same gauge orbit if and only if the corresponding Lie algebra representations are equivalent.
\end{proposition} 

In order to get some interesting gauge field content, one has to couple this noncommutative geometry to the ordinary geometry of a manifold. This is the next noncommutative geometry I will expose.

\subsection[The matrix valued functions algebra]{The matrix valued functions algebra}

In this example, one consider the tensor product of the two algebras $C^\infty(\varM)$ and $M_n(\gC)$ studied before: $\algA = C^\infty(\varM) \otimes M_n(\gC)$. This algebra is the space of smooth matrix valued functions on the smooth paracompact manifold $\varM$ ($\dim \varM = m$).

The derivation-based differential calculus for this algebra was first considered in \cite{DuViKernMado:90b}:
\begin{proposition}[General properties of the differential calculus]
\label{prop-Generalpropertiesofthedifferentialcalculustrivialcase}
One has the following results:
\begin{itemize}
\item $\caZ(\algA) = C^\infty(\varM)$.
\item $\der(\algA) = [\der(C^\infty(\varM))\otimes \bbbone ] \oplus [ C^\infty(\varM) \otimes \der(M_n) ] = \Gamma(\varM) \oplus [C^\infty(\varM) \otimes \ksl_n]$ as Lie algebras and $C^\infty(\varM)$-modules. In the following, we will use the notations: $\kX = X + \adrep_\gamma$ with $X \in \Gamma(\varM)$ and $\gamma \in C^\infty(\varM) \otimes \ksl_n = \algA_0$ (traceless elements in $\algA$).\\
One has the identifications $\Int(\algA) = \algA_0$ and $\Out(\algA) = \Gamma(\varM)$.

\item $\underline{\Omega}^\grast_\der(\algA) = \Omega^\grast_\der(\algA) = \Omega^\grast(\varM) \otimes \Omega^\grast_\der(M_n)$ with the differential $\hd = \dd + \dd'$, where $\dd$ is the de~Rham differential and $\dd'$ is the differential introduced in the previous example (matrix geometry).

\item The noncommutative $1$-form $i\theta$ is defined as $i\theta(X + \adrep_\gamma) = \gamma$. It splits the short exact sequence of Lie algebras and $C^\infty(\varM)$-modules
\begin{equation}
\label{eq-splittingsecderivationstrivialcase}
\xymatrix@1@C=25pt{{\algzero} \ar[r] & {\algA_0} \ar[r] & {\der(\algA)} \ar[r] \ar@/_0.7pc/[l]_-{i\theta}& {\Gamma(\varM)} \ar[r] & {\algzero}}
\end{equation}

\item Noncommutative integration is a well-defined map of differential complexes
\begin{align*}
\int_{\text{n.c.}} : \Omega^\grast_\der(\algA) & \rightarrow \Omega^{\grast - (n^2-1)}(\varM)
&
\int_{\text{n.c.}} \hd \omega &= \dd \int_{\text{n.c.}} \omega
\end{align*}

\end{itemize}
\end{proposition} 

Using a metric $h$ on $\varM$ and the metric $g_{k \ell} = \frac{1}{n} \tr(E_k E_\ell)$ on the matrix part, one can define a metric on $\der(\algA)$ as follows,
\begin{equation*}
\widehat{g}(X+ \adrep_\gamma, Y + \adrep_\eta) = h(X,Y) + \textstyle \frac{1}{\Lambda^2}g(\gamma \eta)
\end{equation*}
where $\Lambda$ is a positive constant which measures the relative ``weight'' of the two ``spaces''. In physical natural units, it has the dimension of a mass.

\bigskip
Consider now the right $\algA$-module $\modM = \algA$. As for the algebra $M_n$, the noncommutative $1$-form $-i\theta$ defines a canonical noncommutative connection by the relation $\widehat{\nabla}^{-i\theta}_\kX a = \kX a - i\theta(\kX) a$ for any $a \in \algA$.

\begin{proposition}[Properties of $\widehat{\nabla}^{-i\theta}$]
For any $a \in \algA$ and $\kX = X + \adrep_\gamma \in \der(\algA)$, one has $\widehat{\nabla}^{-i\theta}_\kX a = X \cdotaction a - a \gamma$.

The curvature of the noncommutative connection $\widehat{\nabla}^{-i\theta}$ is zero.

The gauge transformed connection $\widehat{\nabla}^{-i\theta g}$ by $g \in C^\infty(\varM) \otimes GL_n(\gC)$ is associated to the noncommutative $1$-form $\kX \mapsto -i\theta(\kX) + g^{-1} (X\cdotaction g) = - \gamma + g^{-1} (X\cdotaction g)$.
\end{proposition} 

In \cite{DuViKernMado:90b}, the gauge field content of this noncommutative geometry has been explored. Any noncommutative form splits into parts according to the bigrading of the tensor product $\Omega^\grast(\varM) \otimes \Omega^\grast_\der(M_n)$. Thus, the noncommutative $1$-form defining a connection is a sum of two terms: the first one, in $\Omega^1(\varM) \otimes M_n$, can be interpreted as an ordinary Yang-Mills connection, and the second one, in $C^\infty(\varM) \otimes \Omega^1_\der(M_n)$, has been shown to have some similitudes with Higgs fields. Indeed, the curvature of such a connection is itself a sum of three terms: the first one in $\Omega^2(\varM) \otimes M_n$ is the Yang-Mills curvature of the ``ordinary connection'', the second one in $\Omega^1(\varM) \otimes \Omega^1_\der(M_n)$ is a covariant derivative of the ``Higgs components'', and the third one in $C^\infty(\varM) \otimes \Omega^2_\der(M_n)$ is a second order polynomial expression in the ``Higgs components''. Constructing a natural action with this curvature gives rise to a spontaneous symmetry breaking phenomena in the last term of the curvature, which generates masses for the ``ordinary connection'' through the coupling defined by the second term. 

The models one can construct using this geometry are then natural Yang-Mills-Higgs models. We refer to \cite{DuViKernMado:90b} for more details. This noncommutative geometry has also been used to construct non-abelian generalization of Born-Infeld theories in \cite{Mass:22} and \cite{Mass:26}.

\section{Generalizations of the fundamental examples}

In this section, I would like to generalize the two previous examples in two different directions. 

The first one has been now explored in great details. It consists to replace matrix valued functions on a manifold by sections of a matrix bundle over this manifold. 

The second one consists to take infinite matrices, using some topology to maintain some structural properties.

\subsection[The algebra of endomorphisms of a $SU(n)$-vector bundle]{The algebra of endomorphisms of a $SU(n)$-vector bundle}

The noncommutative geometry described by the algebra $C^\infty(\varM) \otimes M_n(\gC)$ is the trivial situation of some more general geometry described here. This geometry has been studied in \cite{Mass:14}, \cite{Mass:15} and \cite{Mass:25}. A recent review can be found in \cite{Mass:30}, with emphases on the relations between this noncommutative geometry and ordinary geometry of fiber bundles.

Let $\varE$ be a $SU(n)$-vector bundle over the paracompact manifold $\varM$ with fiber $\gC^n$. One can ossociate to it its fiber bundle of endomorphisms, denoted by $\End(\varE)$. Let now $\algA$ be the algebra of sections of $\End(\varE)$. This is the algebra we are interested in. 

In case the vector bundle $\varE = \varM \times \gC^n$ is the trivial fiber bundle, the associated algebra is nothing more than $\algA = C^\infty(\varM)\otimes M_n$. In general, for a non trivial $\varE$, $\algA$ is globally more complicated. Nevertheless, using local trivialisations of $\varE$, the algebra $\algA$ looks locally like $C^\infty(U) \otimes M_n$. 

Here is a summary of the properties of its derivation-based noncommutative geometry, where $\varP$ denote the underlying $SU(n)$-principal fiber bundle to which $\varE$ is associated.
\begin{proposition}[Basic properties]
One has $\caZ(\algA) = C^\infty(\varM)$.

Involution, trace map and determinant ($\tr, \det : \algA \rightarrow C^\infty(\varM)$) are well defined fiberwise. 

Let us define $SU(\algA)$ as the unitaries in $\algA$ of determinant $1$, and $\ksu(\algA)$ as the traceless antihermitean elements. Then $SU(\algA)$ is the gauge group of $\varP$ and $\ksu(\algA)$ is its Lie algebra.
\end{proposition} 

This proposition shows in particular that this noncommutative geometry is strongly related to the gauge geometry defined by $\varP$. From a gauge field perspective, this algebra contains as much as $\varP$, and reveals some great relations between Yang-Mills fields and Higgs fields (as the trivial case defined below already suggests).

\bigskip
Denote by $\rho : \der(\algA) \rightarrow \der(\algA)/\Int(\algA) = \Out(\algA)$ the projection of the short exact sequence \eqref{eq-secderivations}. 
\begin{proposition}[The Lie algebra of derivations of $\algA$ and the differential calculus]
\label{prop-thederivationsodalgA}
One has $\Out(\algA) \simeq \der(C^\infty(\varM)) = \Gamma(\varM)$ and $\rho$ is the restriction of derivations $\kX \in \der(\algA)$ to $\caZ(\algA) = C^\infty(\varM)$. $\Int(\algA)$ is isomorphic to $\algA_0$, the traceless elements in $\algA$.

The short exact sequence of Lie algebras and $C^\infty(\varM)$-modules of derivations looks like
\begin{equation*}
\xymatrix@R=0pt@C=15pt{ 
{\algzero} \ar[r] & {\Int(\algA)} \ar[r] & {\der(\algA)} \ar[r]^-{\rho} & {\Gamma(\varM)} \ar[r] & {\algzero}\\
           &                    &  \kX        \ar@{|->}[r]   & X       &
}
\end{equation*}
Real inner derivations are given by the $\adrep_\xi$ with $\xi \in \ksu(\algA)$.

One has the identification
\begin{equation*}
\underline{\Omega}^\grast_\der(\algA) = \Omega^\grast_\der(\algA)
\end{equation*}

\end{proposition} 

In the trivial case, one has a splitting of this short exact sequence of Lie algebras and $C^\infty(\varM)$-modules. This is no longer true in the topologically non trivial case. Moreover, the noncommutative $1$-form $i\theta$ is no more defined here, but one can define a map of $C^\infty(\varM)$-modules:
\begin{align*}
i\theta : \Int(\algA) &\rightarrow \algA_0
&
\adrep_\gamma &\mapsto \textstyle\gamma - \frac{1}{n}\tr(\gamma) \bbbone
\end{align*}

The Cartan operation of the Lie algebra $\Int(\algA)$ plays an essential role in the study of ordinary connections on $\varE$ and their relations to the noncommutative geometry of $\algA$. This is explained in great details in \cite{Mass:30}.

\bigskip
The main result connecting the ordinary gauge theory on $\varE$ and the noncommutative geometry of $\algA$ is contained in the following construction. To any connection $\nabla^\varE$ on the vector bundle $\varE$ one can associate two connections, $\nabla^{\varE^\ast}$ on $\varE^\ast$ (the dual vector bundle of $\varE$) and $\nabla$ on $\End(\varE)$, by the relations
\begin{align*}
X\cdotaction \langle \varphi, s \rangle &= \langle \nabla_X^{\varE^\ast} \varphi,  s \rangle + \langle \varphi, \nabla_X^\varE s \rangle
&
\nabla_X (\varphi \otimes s) &= (\nabla_X^{\varE^\ast} \varphi) \otimes s + \varphi \otimes (\nabla_X^\varE s)
\end{align*}
with $X \in \Gamma(\varM)$, $\varphi \in \Gamma(\varE^\ast)$ and $s \in \Gamma(\varE)$, where we use the usual identification $\End(\varE) \simeq \varE^\ast \otimes \varE$.

From now on, we will use the notation $X = \rho(\kX) \in \Gamma(\varM)$ for any $\kX \in \der(\algA)$.

\begin{proposition}[The noncommutative $1$-form $\alpha$]
\label{prop-Thenoncommutative1formalpha}
For any $X \in \Gamma(\varM)$, $\nabla_X$ is a derivation of $\algA$. 

For any $\kX \in \der(\algA)$, the difference $\kX - \nabla_X$ is an inner derivation, so that one can define the map $\kX \mapsto \alpha(\kX)= -i\theta(\kX - \nabla_X)$. By construction, $\alpha$ is a noncommutative $1$-form $\alpha \in\Omega^1_\der(\algA)$ which induces the decomposition
\begin{equation*}
\kX = \nabla_X - \adrep_{\alpha(\kX)}
\end{equation*}

For any $\gamma \in \algA_0$, one has $\alpha(\adrep_\gamma) = -\gamma$; for any $\kX \in \der(\algA)$, one has $\tr \alpha(\kX)= 0$; and for any $\kX \in \der_\gR(\algA)$, one has $\alpha(\kX)^\ast + \alpha(\kX) = 0$.
\end{proposition} 

This show in particular that $X\mapsto \nabla_X$ is a splitting as $C^\infty(\varM)$-modules of the short exact sequence
\begin{equation}
\label{eq-splittingofshortexactsequenceofderivations}
\xymatrix@1@C=25pt{{\algzero} \ar[r] & {\algA_0} \ar[r] & {\der(\algA)} \ar[r] & {\Gamma(\varM)} \ar[r] \ar@/_0.7pc/[l]_-{\nabla} & {\algzero}}
\end{equation}
The obstruction to be a splitting of Lie algebras is nothing but the curvature of $\nabla$: $R(X,Y) = [\nabla_X, \nabla_Y] - \nabla_{[X,Y]}$.

Moreover, $\alpha$ can be considered as an extension of $-i \theta$ to all $\der(\algA)$. The main result is then:
\begin{proposition}[Ordinary connections as noncommutative $1$-forms]
\label{prop-ordinaryconnectionsandnoncommutativeforms}
The map $\nabla^\varE \mapsto \alpha$ is an isomorphism between the affine spaces of $SU(n)$-connections on $\varE$ and the traceless antihermitean noncommutative $1$-forms on $\algA$ such that $\alpha(\adrep_\gamma) = -\gamma$.

The noncommutative $2$-form $(\kX, \kY) \mapsto \Omega(\kX, \kY) = \hd\alpha(\kX, \kY) + [\alpha(\kX), \alpha(\kY) ]$ depends only on the projections $X$ and $Y$ of $\kX$ and $\kY$. This means that it is a horizontal noncommutative $2$-form for the operation of $\Int(\algA)$ on $\Omega^\grast_\der(\algA)$.

The curvature $R^\varE$ of $\nabla^\varE$ identifies with the horizontal noncommutative $2$-form $\Omega$.

The noncommutative $1$-form $\alpha^u$ corresponding to the gauge transformed connection $\nabla^{\varE u}$ induced by the gauge transformation $u \in SU(\algA)$ is given by 
\begin{equation*}
\alpha^u = u^\ast \alpha u + u^\ast \hd u
\end{equation*}

\end{proposition} 

As a consequence, infinitesimal gauge transformations on connections on $\varE$ express themselves as Lie derivatives by real inner derivations on $\alpha$.

\bigskip
Let us consider now the right $\algA$-module $\modM=\algA$. Then the noncommutative $1$-form $\alpha$, associated to an ordinary connection $\nabla^\varE$ on $\varE$, defines a noncommutative connection $\widehat{\nabla}^\alpha$ on $\modM$ by the relation $\widehat{\nabla}^\alpha_\kX a = \kX a + \alpha(\kX) a$.

\begin{proposition}[The noncommutative connection associated to $\alpha$]
One has
\begin{equation*}
\widehat{\nabla}^\alpha_\kX a = \nabla_X a + a \alpha(\kX)
\end{equation*}
In particular, for any $X \in \Gamma(\varM)$, one has $\widehat{\nabla}^\alpha_{\nabla_X} a = \nabla_X a$.

This noncommutative connection $\widehat{\nabla}^\alpha$ is compatible with the canonical Hermitean structure.

The curvature of $\widehat{\nabla}^\alpha$ is $\hR^\alpha(\kX, \kY) = R^\varE(X,Y)$.

A gauge transformation induced by $u \in SU(\algA)$ on the connection $\nabla^\varE$ induces a (noncommutative) gauge transformation on $\widehat{\nabla}^\alpha$.
\end{proposition} 

This proposition then implies the following important practical result:
\begin{theorem}[Ordinary connections as noncommutative connections]
The space of noncommutative connections on the right $\algA$-module $\algA$ compatible with the Hermitean structure $(a,b) \mapsto a^\ast b$ contains the space of ordinary $SU(n)$-connections on $\varE$. 

This inclusion is compatible with the corresponding definitions of curvature and gauge transformations.
\end{theorem} 

This means that one can forget about ordinary $SU(n)$-connections, and replace this notion by the more general notion of noncommutative connections on the right $\algA$-module $\algA$. From a physical point of view, the degrees of freedom one gets in this embedding correspond to some fields that it is possible to interpret as Higgs fields.

Using a definition of ``noncommutative quotient manifolds'' introduced in \cite{Mass:07} in the context of the derivation-based noncommutative geometry, the noncommutative geometry of $\algA$ can be related through some Cartan operations of the Lie algebra $\ksu(n)$ to the (ordinary) geometry of the underlying $SU(n)$-principal fiber bundle $\varP$ (see \cite{Mass:30} for a complete review). Moreover, using such a Cartan operation, the notion of noncommutative invariant connections has been defined and studied in \cite{Mass:25}.

\subsection[The Moyal algebra]{The Moyal algebra}

In order to simplify the exposition given here, I will restrict to the $2$-dimensional case of the Moyal algebra, which means that I will only consider the deformation of the algebra of functions on the plane $\gR^2$. 

There are various possibilities to introduce the so-called Moyal algebra, and there are indeed many such algebras. In \cite{GraciaBondia:1987kw}, \cite{Varilly:1988jk} and \cite{Gayral:2003dm}, some possibles definitions of the Moyal algebras are given in details, and many considerations about these constructions and their topological aspects are exposed. 

Here, I will consider one particular construction, which has been used in many studies of non commutative field theories on ``Moyal spaces'' (see \cite{RIVASSEAU:2007:HAL-00165686:1} and \cite{WALLET:2007:HAL-00170965:1} for recent reviews, and references therein). In \cite{Gayral:2003dm}, it is called the ``Moyal multiplier algebras'' for reasons that will be clear in a while. Let us summarize the definition which is used in the following.

Denote by $\caS(\gR^2)$ the space of complex-valued Schwartz functions on the plane $\gR^2$, and by $\caS'(\gR^2)$ the space of associated tempered distributions. Let $\Theta = \theta \spmatrix{0 & -1 \\ 1 & 0}$ be an antisymmetric matrix, with $\theta \in \gR$, $\theta \neq 0$, a deformation parameter. Then one can define the Moyal-Groenenwald product $\caS(\gR^2) \times \caS(\gR^2) \rightarrow \caS(\gR^2)$ by the integral formula
\begin{equation*}
(f \moyast g)(x)=\frac{1}{(\pi\theta)^2}\int d^2 y d^2 z\ f(x+y) g(x+z) e^{-i2 y \Theta^{-1} z} 
\end{equation*}

This product can be extended to give a left and a right module structures on $\caS'(\gR^2)$ by the relations
\begin{align*}
\caS(\gR^2) \times \caS'(\gR^2) &\rightarrow \caS'(\gR^2)
&
\caS'(\gR^2) \times \caS(\gR^2) &\rightarrow \caS'(\gR^2)
\\
\langle f \moyast T, g \rangle &= \langle T, g \moyast f \rangle
&
\langle T \moyast f, g \rangle &= \langle T, f \moyast g \rangle
\end{align*}
where $\langle T , f \rangle$ is the coupling between $\caS'(\gR^2)$ and $\caS(\gR^2)$. The smoothening property of the Moyal product ensures that $f \moyast T$ and $T \moyast f$ are smooth functions.

One can then define the left and right multiplier spaces by
\begin{align*}
\caL &= \{ T \in \caS'(\gR^2) \ / \ f \moyast T \in \caS(\gR^2), \forall f \in \caS(\gR^2) \}\\
\caR &= \{ T \in \caS'(\gR^2) \ / \ T \moyast f \in \caS(\gR^2), \forall f \in \caS(\gR^2) \}
\end{align*}
This leads to the following definition of the Moyal algebra:
\begin{definition}[Moyal algebra]
The Moyal algebra is $\algA_\Theta = \caL \cap \caR$.
\end{definition} 

By construction, the Moyal algebra contains $\caS(\gR^2)$ as an ideal. But, as the following proposition shows, it contains also other particular functions which do not vanish at infinity:
\begin{proposition}[Polynomial functions and the Moyal product]
Polynomial functions are in $\algA_\Theta$ (in particular $\algA_\Theta$ is a unital algebra), and the Moyal product of two polynomial functions is given by the following expression, which is in fact a finite sum,
\begin{multline*}
(P\moyast Q)(x) = P(x) \cdotaction Q(x) + \\
 \sum_{n=1}^\infty \frac{1}{n!} \left(\frac{i}{2} \Theta^{\mu_1\nu_1} \frac{\partial^2}{\partial x^{\mu_1} \partial y^{\nu_1}}\right) \cdots \left(\frac{i}{2} \Theta^{\mu_n\nu_n} \frac{\partial^2}{\partial x^{\mu_n} \partial y^{\nu_n}}\right) P(x) Q(y) {}_{\vert_{x=y}}
\end{multline*}
In particular, this leads to the (well known) commutators: $[x^\mu,x^\nu]_\moyast = i\Theta^{\mu\nu}$.
\end{proposition}

\begin{proposition}[Center and derivations of the Moyal algebra]
$\caZ(\algA_\Theta) = \gC$ and $\der(\algA_\Theta) = \Int(\algA_\Theta)$.
\end{proposition} 

For instance, one has $\partial_\mu a = [ -i\Theta^{-1}_{\mu\nu} x^\nu, a]_\moyast$, for any $a \in \algA_\Theta$.

\begin{remark}[The infinite dimensional vector space $\der(\algA_\Theta)$]
As a mere consequence of the previous result, $\der(\algA_\Theta)$ is an \emph{infinite dimensional} vector space, which means in particular that it is not a finite projective module over the center. This situation is quite different to the ones encountered so far in the previous exemples exposed before, the algebras $C^\infty(\varM)$, $M_n(\gC)$, $C^\infty(\varM) \otimes M_n(\gC)$ and the endomorphism algebra of a $SU(n)$-vector bundle. 

If one defines the derivation-based differential calculus directly from this infinite dimensional Lie algebra, this would imply strange and inconvenient manifestations for gauge fields theories. Indeed, using the right module $\algA_\Theta$ for such a gauge theory, any connection would have an infinite number of field components $A_a \in \algA_\Theta$, one for each element $X_a$ of a basis of $\der(\algA_\Theta)$. From this, it follows that it is highly desirable to concentrate on a finite dimensional Lie sub algebra of $\der(\algA_\Theta)$.

\end{remark}

\begin{remark}[The algebras of commutation relations]
An other situation where the dimension of the Lie algebra of derivations is infinite is the algebra generated (algebraically) by the commutation relations $[p,q] = i$. This algebra is often called the Heisenberg algebra.

It has been equipped with a symplectic form by M.~Dubois-Violette in~\cite{DuVi:91}; this permits one to look at commutators as Poisson brackets.

Notice that this algebra is not the Moyal algebra, even if some confusion can been introduced if one notices that these two algebras are generated by the same commutation relations. Indeed, the Heisenberg algebra is generated algebraically by these relations, and it admits the product inherited by them. On the contrary, the Moyal algebra is a deformation algebra, and it admits a more complicated and may be a more confusing structure. The Moyal product, denoted here by $\moyast$, has been construted to be defined everywhere on $\algA_\Theta$. But the ordinary product of functions, inherited from the construction through the ideal $\caS(\gR^2)$, and defined on a larger class of elements (as polynomial functions for instance), is largely used and of interest to study this geometry. Then, in the Heisenberg algebra, one has $pq \neq qp$, but in the Moyal algebra, one has $x^\mu x^\nu = x^\nu x^\mu$ and $x^\mu \moyast x^\nu \neq x^\nu \moyast x^\mu$.

To add to the confusion, some authors often introduce the relation 
\begin{equation*}
\int f \moyast g = \int fg
\end{equation*}
to simplify computations. This has to be considered very delicately. Indeed, it makes sense only for the subclass of elements on which this ordinary product is defined, which certainly contains a large class of functions, but definitively not some pure distributions!
\end{remark}

In \cite{GraciaBondia:1987kw}, it is shown, using some results in \cite{Simo:71a}, that the space $\caS(\gR^2)$ admits a double indexed basis $f_{mn}$ which identifies $\caS(\gR^2)$ with double indexed rapidly decreasing sequences $(s_{mn})$ through the decomposition $s = \sum_{m,n \geq 0} s_{mn} f_{mn}$. Defining the family of seminorms
\begin{equation*}
r_k(s) = \left[ \sum_{m,n\geq 0} (2m+1)^{2k} (2n+1)^{2k} | s_{mn} |^2  \right]^{1/2}
\end{equation*}
for such a decomposition, one gets that $s \in \caS(\gR^2)$ if and only if $r_k(s)$ is finite for every $k\geq 0$.

In \cite{Varilly:1988jk}, it is then shown that $\algA_\Theta$ is the algebra of infinite dimensional matrices $a = (a_{mn})$ for which $as = (\sum_{n\geq 0} a_{mn} s_{np})$ and $sa = (\sum_{n\geq 0} s_{mn} a_{np})$ are in $\caS(\gR^2)$ for any $s \in \caS(\gR^2)$. The topology of this space of left and right operators on $\caS(\gR^2)$ is the operator topology. In this identification, the Moyal product $\moyast$ is exactly the (infinite dimensional) matrix product.

Using this identification, one can now look at the Moyal algebra as a completion of the algebra $M_\infty = \varinjlim M_n$ into a locally convex unital algebra, so that this algebra is a generalisation of the example of the (finite dimensional) matrix algebra. As we will see, it shares indeed a lot of properties with this algebra.

\begin{remark}[Other ``Moyal algebras'']
\label{remark-othermoyalalgebras}
In \cite{Gayral:2003dm}, the Moyal algebra is defined to be the unital Fréchet pre-$C^\ast$-algebra of smooth functions on $\gR^2$ bounded together with all derivatives, hereafter denoted by $\algB$. Obviously, this algebra is not the one considered here ($\algB$ does not contain polynomial functions for instance) but $\algB \subset \algA_\Theta$. As a consequence, the derivations $a \mapsto \partial_\mu a = [ -i\Theta^{-1}_{\mu\nu} x^\nu, a]_\moyast$ are not inner derivations anymore.

This difference needs to be commented. In \cite{Gayral:2003dm}, an argumentation is provided to disregard $\algA_\Theta$ as a too vast algebra to be of ``practical use''. This does not seem to correspond to what the bibliography references indicates. In fact, the very reason why the authors of \cite{Gayral:2003dm} needed to consider only $\algB$, and not the full algebra $\algA_\Theta$, was to get a natural representation of their algebra on the space of functions $f \in L^2(\gR^2)$ by left Moyal multiplication. This would be impossible with the big algebra $\algA_\Theta$ considered here. This representation is used to construct the spectral triple which they wanted to show the existence\dots

Let me also point out that a very brief allusion is made to the fact that the theory of finite projective modules over $\algA_\Theta$ is not suitable to consider noncommutative vector bundle.
Beside the fact that the authors only mentioned finite projective modules in one of the axioms of the spectral triple they constructed, this point would be important in general. But with the particular algebra $\algA_\Theta$, the existence of ``interesting'' finite projective modules need not be worried about, at least from a gauge fields theories point of view. Finite projective modules are very convenient in general if one wants to consider noncommutative connections, because this property ensures the existence of such noncommutative connections. This is a sufficient condition, but not a necessary one. Indeed, as will be clear in a while, for the differential calculus considered in the following, the existence of at least one noncommutative connection is ensured for every module over $\algA_\Theta$! This strong result is a trivial consequence of the existence of a particular $1$-form in the differential calculus introduced below.
\end{remark}

\begin{remark}[Modules and ``matter fields'']
Let me make some comments about modules over $\algA_\Theta$ and their relations to matter fields in physics. In ordinary geometry a module over an algebra (let me forget to mention for a while some properties like finite projective, hermitien, \textit{etc}) is the space of matter fields, as sections of a vector bundle, and gauge fields are the components of some connections acting on this vector bundle. This situation is transposed in noncommutative geometry in the same algebraic terms. For instance, in the noncommutative version of the standard model of elementary particles by Chamseddine, Connes and Marcolli \cite{Chamseddine:2006ep}, matter fields are elements of the Hilbert space which support a bimodule structure of the algebra of the spectral triple, and the Dirac operator of the triple is then a first order operator for this bimodule structure (see \cite{Mass:08} for general properties of first order operators in bimodules). 

This situation is abviously also at the heart of gauge fields theories on Moyal spaces (see \cite{DEGOURSAC:2007:HAL-00135917:1}, \cite{Grosse:2006hh}, \cite{Grosse:2007dm}, \cite{deGoursac:2008rb} and \cite{WALLET:2007:HAL-00170965:1} and references therein). But in this situation, the fact that $\caS(\gR^2)$ is an ideal of $\algA_\Theta$ implies that it can be used as a right $\algA_\Theta$-module. Then matter fields are rapidly decreasing functions on $\gR^2$, but gauge fields do not need to: they can grow like polynomial functions! And indeed, there are interesting connections on the Moyal plane with this property.
\end{remark}

As noted before, the Lie algebra of derivations of $\algA_\Theta$ is too big to be useful in gauge fields theories, so that one has to choose a smaller Lie subalgebra, with finite dimension (in order to get a finite number of gauge potentials).

The natural choice that has been assumed in a lot of considerations on fields theories on this noncommutative space is the one consisting of the Lie algebra whose basis is $\{\partial_\mu\}_{\mu=1,2}$. This choice is not clearly expressed in the literature, because it is the most natural one. In that case, introducing models via noncommutative connections on the right module $\algA_\Theta$ itself, there are two gauge potentials $\{A_\mu\}_{\mu=1,2}$, defined by $\widehat{\nabla}_{\partial_\mu} \bbbone = A_\mu$. Models using this structure have been considered in \cite{DEGOURSAC:2007:HAL-00135917:1}, \cite{Grosse:2006hh}, \cite{Grosse:2007dm}, \cite{deGoursac:2008rb} and \cite{WALLET:2007:HAL-00170965:1} (for a review). Two mains problems remain to be solved in this context: to find renormalizable gauge field theories and to describe (some of) their vacuum configurations. 
Owing to the fact that $\partial_\mu a = [ -i\Theta^{-1}_{\mu\nu} x^\nu, a]_\moyast$, the derivations used in this differential calculus are associated to polynomial functions of degree less than or equal to $1$.

In \cite{Mass:32}, we proposed to consider a bigger Lie algebra, consisting of derivations associated to polynomial functions of degree less than or equal to $2$. This Lie subalgebra is isomorphic to $\kisp(2,\gR)^\gC$, the complexified Lie algebra of the inhomogeneous symplectic Lie algebra on $\gR^2$. This choice is based on the following facts, which make this Lie algebra very convenient:
\begin{proposition}
Let $P$ and $Q$  be polynomial functions of degree less than or equal to $2$. 

One has
\begin{equation*}
[P,Q]_\moyast = \{ P, Q\}_\text{PB}
\end{equation*}
where $\{ P, Q\}_\text{PB}$ is the Poisson bracket associated to the Poisson structure defined by the antisymmetric matrix $\Theta$.

The inner derivation induced by $P$ is an ordinary derivation on the maximal commutative subalgebra in $\algA_\Theta$ containing $\caS(\gR^2)$ and the polynomial functions.
\end{proposition}

In particular, any derivation in $\kisp(2,\gR)^\gC$ can be considered as the infinitesimal version of an area preserving automorphism of $\gR^2$. This is the maximal Lie subalgebra of $\der(\algA_\Theta)$ to have this property. This means that the ``directions'' along the derivations have some direct geometrical interpretations: the two derivations $\partial_\mu$, $\mu=1,2$, are associated to ordinary translations, so that they will be called ``spatial'' directions in the following, while the three others are associated to symplectic rotations.

We will denote by $(\Omega_\kisp^\grast(\algA_\Theta), d)$ the derivation-based differential calculus constructed on this Lie subalgebra.

\begin{remark}
This Lie subalgebra, as well as its properties which are given in the previous proposition, has already been described in \cite{MarmVitaZamp:06a} in a similar but slightly different context. The main object of this paper is to generate some natural (associative) subalgebras of the Moyal algebra on $\gR^4$ using some three dimensional Lie subalgebras of $\kisp(2,\gR)$ as constraints, and to recover, in a commutative limit, the algebra of smooth functions on $\gR^3$. This construction relies on ideas developed in \cite{Madore:1991bw} on the fuzzy sphere, where a Lie subalgebra of dimension $3$  of the matrix algebra $M_n(\gC)$ (the usual $n$-dimensionnal representation of $\ksu(2)$), is used to construct noncommutative ``spaces'' which are approximations of the $2$-dimensional sphere. In \cite{MarmVitaZamp:06a}, no considerations are made about field theories on the Moyal plane equipped with this differential calculus. 
\end{remark}

In the study of gauge theories on the algebra $C^\infty(\varM) \otimes M_n(\gC)$, the gauge potentials in the noncommutative directions were identified with Higgs fields. What about the present situation? This has been explored in \cite{Mass:32}. Here are the main results.

\begin{proposition}
There exists a canonical noncommutative $1$-form $\eta \in \Omega_\kisp^1(\algA_\Theta)$ such that $da = [\eta, a]$.

This noncommutative $1$-form induces a canonical connection on any right $\algA_\Theta$-module, whose curvature depends only on the matrix $\Theta$.
\end{proposition}

The canonical noncommutative $1$-form $\eta$ is defined by $\eta(\adrep_P) = P_0$ where, for any polynomial function $P$ of degree less than or equal to $2$, $P_0$ is the polynomial function $P$ from which we remove the constant part (which is in the center of $\algA_\Theta$). For instance, one gets $\eta(\partial_\mu) = -i\Theta^{-1}_{\mu\nu} x^\nu$ for the spatial directions.

For any right $\algA_\Theta$-module $\modM$, the canonical noncommutative connection is defined by the relation $\widehat{\nabla}_\kX m = - m \eta(\kX)$ for any $m \in \modM$ and $\kX \in \kisp(2,\gR)^\gC$. Its curvature is $\hR(\kX, \kY) m = m \left( \eta([\kX, \kY]) - [\eta(\kX), \eta(\kY)] \right)$. The ``curvature''  $\eta([\kX, \kY]) - [\eta(\kX), \eta(\kY)]$ is non zero only in the spatial directions, where it takes values proportional to $\Theta^{-1}_{\mu\nu}$.

This canonical noncommutative $1$-form $\eta$ has to be compared with the canonical noncommutative $1$-form $i\theta \in \Omega^1_\der(M_n)$ of Proposition~\ref{prop-differentialcalculusmatrixalgebra}. But, contrary to $i\theta$, $\eta$ cannot be adjusted to be a morphism of Lie algebras, so that the curvature of the associated canonical connection is non zero.

The interpretation of the ``noncommutative directions'' of the gauge potentials as Higgs fields for the algebra $C^\infty(\varM) \otimes M_n(\gC)$ relies heavily on the fact that the canonical connection defined by $i\theta$ is of zero curvature. In the present situation, as noted before, in the directions of the symplectic rotations the curvature of the canonical connection is zero. This fact has been used in \cite{Mass:32} to show that, indeed, the gauge potentials in the directions of the symplectic rotations look very much like Higgs fields. In this situation, the symplectic rotations are the ``inner'' derivations for this geometry, and the spatial translations are the ``outer'' ones (even if all derivations are inner from an algebraic point of view).

This interpretation is also sustained by the fact that for the algebra $\algB$ introduced in Remark~\ref{remark-othermoyalalgebras}, the derivations $\partial_\mu$ are outer derivations, and the derivations in the directions of the symplectic rotations are not defined at all. From this point of view, $\algB$ plays, with respect to $\algA_\Theta$, the same role that the algebra $C^\infty(\varM)$ plays with respect to $C^\infty(\varM) \otimes M_n(\gC)$ (or $\End(\varE)$ for a $SU(n)$-vector bundle $\varE$ over $\varM$).

\bibliographystyle{unsrt} 
\bibliography{biblio-articles-perso,biblio-livre,biblio-articles}

\end{document}